\documentclass[bibyear]{aa} 

\usepackage{graphicx}
\usepackage{multirow}
\usepackage{txfonts}
\usepackage[usenames]{color}

\begin{document} 

    \titlerunning{Estimating the mass-to-distance ratio for AGN black holes using general relativity}
    \authorrunning{A. Gonz\'alez-Ju\'arez et al.}
    \title{Estimating the mass-to-distance ratio for a set of megamaser AGN black holes by employing a general relativistic method}


   \author{A. Gonz\'alez-Ju\'arez,
          \inst{1}
          M. Momennia,
          \inst{1,2}
          A. Villalobos-Ram\'irez,
          \inst{1}
          \and
          A. Herrera-Aguilar\inst{1}\fnmsep\thanks{Just to show the usage
          of the elements in the author field}
          }

   \institute{Instituto de F\'\i{}sica Luis Rivera Terrazas (IFUAP), Benem\'{e}rita Universidad Aut\'{o}noma de Puebla, Puebla 72570, M\'exico\\
         \and
             Instituto de F\'{\i}sica y Matem\'{a}ticas, Universidad Michoacana de San Nicol\'as de Hidalgo, Michoac\'{a}n 58040, M\'exico\\
             }

   \date{Received Month XX, XXXX; accepted Month XX, XXXX}

 
  \abstract
   {Motivated by the recent achievements of a full general relativistic method in determining black hole (BH) parameters, we continue to estimate the mass-to-distance ratio of the supermassive BHs hosted at the core of the active galactic nuclei (AGNs) of the megamaser galaxies NGC 1320, NGC 1194, NGC 5495, and Mrk 1029.}
   {Our aim is to study the properties of super massive BHs at the centers of the selected AGNs by using a full general relativistic method that allows us to address the potential detection of relativistic effects within such astrophysical systems.}
   {In order to perform statistical estimations with publicly available observational data, we used a general relativistic model that describes BH rotation curves and further employed a Bayesian fitting method.}
  {We estimated the mass-to-distance ratio of the aforementioned BHs, their  
  position and the recessional redshifts of the host galaxies produced by both peculiar motion and cosmological expansion of the Universe.
   Finally, we calculated the gravitational redshift of the closest maser to the BH for each AGN. This gravitational redshift is a general relativistic effect produced by the gravitational field of the BH properly included in the modelling.}
   {}


   \maketitle

\section{Introduction}
Currently, one of the most active research areas in modern astrophysics is the search for observational evidence to support the existence of black holes (BHs) in nature. Although BHs were first found as mathematical solutions to Einstein field equations, several phenomena could potentially lead to the formation of a BH. As the most promising scenarios, we can refer to the collapse of a massive star to form a stellar-mass BH (see Mirabel \cite{Mirabel}), processes in galactic dynamics to create supermassive black holes (SMBHs) in the center of most galaxies (Kormendy \& Richstone \cite{Kormendy}), and fluctuations in the early Universe that could generate primordial BHs (see Hawking \cite{Hawking} and Carr \cite{Carr}).

The discovery of quasars and AGNs gave rise to the idea that SMBHs are located at the core of many galaxies. The reason is that only SMBHs can compellingly explain the production of huge rates of luminous radiation within small compact volumes. A viable explanation of this phenomenon is the conversion of gravitational energy into radiation in their potential well, rendering a long-term stable central mass (see, for instance, Celotti et al. \cite{Review}, and references therein). A tighter constraint on the compactness of the central object comes from the highly energetic X-ray radiation.
This X-ray electromagnetic radiation together with the line emission from gas particles
with velocities up to thousands of kilometers per second can be satisfactorily explained by the presence of a central SMBH in these galaxies (Celotti et al. \cite{Review}).

In the past decades, two groups of astronomers (the Galactic Center teams at MPE and UCLA) have been tracking the orbital motion of a set of stars around the gravitational center of our Galaxy -- Genzel \& Townes (\cite{Genzel87}); Genzel et al. (\cite{Genzel96}); Ghez et al. (\cite{Ghez98}), (\cite{Ghez05}). The detected velocities and accelerations of these stars provided strong evidence for the existence of a central source, called Sagittarius A* (Sgr A*), that is a SMBH with mass of $\approx 4.0 \times 10^{6} M_{\odot}$ (see Do et al. \cite{do2019}, Gravity Collaboration et al. \cite{GRAVITY2019}, \cite{GRAVITY2020}). In these works, the putative BH mass in the center of the Milky Way has been estimated based on Newtonian models with relativistic corrections.

Furthermore, the two SMBHs at the core of M87 and the Milky Way galaxies have been studied by the Event Horizon Telescope (EHT) Collaboration with the aim of capturing their shadow image. In this regard, the EHT Collaboration revealed the first images of the shadows of these SMBHs -- Event Horizon Telescope Collaboration et al. (\cite{EHTI}), (\cite{EHTIV}), (\cite{EHTVI}), (\cite{EHTSGRII}a), (\cite{EHTSGRIV}b). Hence, investigating this phenomenon has attracted much attention recently.

However, it is worth noting that in order to study the putative SMBHs in some of the various aforementioned AGNs, a full general relativistic method is needed, and this shows the importance of the present study. This relativistic method has allowed us to address the potential detection of relativistic effects within such astrophysical systems.

The Megamaser Cosmology Project (MCP)\footnote{%
\url{https://safe.nrao.edu/wiki/bin/view/Main/MegamaserCosmologyProject}}
of the National Radio Astronomy Observatory is a collaboration with the Cosmic Microwave Background project from the Wilkinson Microwave Anisotropy Probe and Planck missions. The primary aim of the MCP is to determine the Hubble constant from galaxies in the Hubble flow whose disks contain water megamasers, Reid et al. (\cite{MCPI}). Moreover, they also estimate the distance to these galaxies and the BH mass hosted in their center.

Water megamasers are water vapor clouds in AGNs emitting at 22 GHz by stimulated emission, moving in nearly circular orbits, and forming thin disk structures mostly observable (almost) edge-on. In the MCP, the authors have used Keplerian motion with relativistic corrections to fit
the trajectories of the masers in the BH accretion disks for 24 different galaxies reported in Reid et al. (\cite{MCPI}); Kuo et al. (\cite{MCPIII}); Kuo et al. (\cite{MCPV}); Gao et al. (\cite{MCPIX}), Zhao et al. (\cite{MCPX}); Kuo et al. (\cite{MCPXII}).

On the other hand, a general relativistic method has been implemented to estimate the mass-to-distance ratio $M/D$ ($D$ is the distance from the
observer to the BH) of the BH hosted at the core of NGC 4258 in 
Nucamendi et al. (\cite{Nucamendi21}). This relativistic model takes into account a system of probe particles that circularly orbit a Schwarzschild BH in the equatorial plane. In this scenario, the observational quantities required to perform the estimations are the frequency shifts experienced by the photons emitted by test particles (in this case, by the megamasers) and their positions (see Herrera-Aguilar \& Nucamendi \cite{Herrera15} and Banerjee et al. \cite{Banerjee22} for the full description of a rotating BH model). Thus, as a result, in Nucamendi et al. (\cite{Nucamendi21}), the $M/D$ ratio of the SMBH of NGC 4258 was estimated to be $M/D = (0.5326\pm 0.0002) \times10^{7} M_{\odot}~{Mpc}^{-1}$. Also, the authors provided the computation of general and special relativistic effects: the gravitational redshift produced by the BH mass and the special relativistic boost that mimics the peculiar motion of the host galaxy from Earth.

As another application of this general relativistic model, we can refer to Villalobos-Ram\'irez et al. (\cite{Artemisa22}), where the $M/D$ ratio of the galaxy TXS-2226-184 was estimated for the first time. Furthermore, the authors quantified the BH mass by using the distance reported in Surcis et al. (\cite{txs2226}) and found that the result is compatible with the mass obtained from the mass-luminosity correlation. By comparing both BH mass results, the accuracy of the general relativistic model is one order of magnitude better than that of the mass-luminosity correlation.

More recently, the $M/D$ ratio has been estimated for ten more galaxies of the MCP systems in Villaraos et al. (\cite{Deby1}). In these water megamaser systems, it is possible to identify three groups. The first group represents frequency-shifted photons close to the systemic redshift of the host galaxy, whereas the other two groups are 
highly frequency-shifted photons coming from regions of the disk tangent to the midline on either side of the center (see, e.g., Lo \cite{KYLo05} and Henkel et al. \cite{Henkel05}).
They found that the velocity corresponding to the gravitational redshift of the closest maser to each BH is about $1-6~$ km~s$^{-1}$ and that BH masses range from $10^{6}$ to $10^{7}M_{\odot }$, a mass domain associated with supermassive BHs.%

In the present study, we apply the general relativistic method and a
Bayesian statistical treatment to investigate four more megamaser systems
among the remaining 14 MCP galaxies. For NGC 1320, we estimated four parameters, including the $M/D$ ratio, 
the BH position along the $x$- and $y$-axes (see Sect. 3.5), and the recession velocity $%
v_{rec}$. Moreover, for the galaxies NGC 1194, NGC 5495, and Mrk 1029, we estimated three parameters, including the $M/D$ ratio, the BH position along the $x$-axis (see Secs. 3.2-4), and $v_{rec}$.
We did not fit the rest of the galaxies because of some
particular features that they display. For example, the case of J1346+5228 has only a few reported masers, one redshifted and one
blueshifted maser, leading to poor statistical results. Redshifted and blueshifted maser lines were not detected in NGC 2824 and J0350-0127. Therefore, since the model is based on the observations of these lines, we could not perform an estimation.
Observations of NGC 6926 showed an extra group of redshifted megamasers whose treatment goes beyond the scope of this
work. Finally, UGC 6093 and Mrk 1 are two galaxies that have been reported with position uncertainties of the same order of those of their angular distance, which led to imprecise results.

We organize this paper as follows: In the next section, we briefly review the general relativistic model implemented in this work. In Sect. \ref{sec:bayes}, the Bayesian approach applied to the model of
section \ref{sec:GR}
is exposed, and the results of the statistical fit are presented. Finally, we discuss the results in Sect. \ref{sec:conclusions}
. 
\section{General relativistic model}
\label{sec:GR}
This BH rotation curve model is based on a general relativistic formalism that was developed in Herrera-Aguilar and Nucamendi (\cite{Herrera15}) and Banerjee et al. (\cite{Banerjee22}) to analytically express the mass and spin parameters of the rotating Kerr BH in terms of observational redshift and blueshift of photons emitted by massive particles orbiting the BH and their orbital parameters. This scenario has been employed in Nucamendi et al. (\cite{Nucamendi21}) for a static and spherically symmetric Schwarzschild BH of the form
\begin{equation}
ds^{2}=\dfrac{dr^{2}}{f}+r^{2}(d\theta ^{2}+\sin {\theta }^{2}d\varphi
^{2})-fdt^{2},\quad \quad f=1-\frac{2m}{r},  \label{schw}
\end{equation}%
where $m$\ is the total mass of the BH in $G=1=c$ units.
In this case, the test particles of the model are water megamasers circularly orbiting a central static BH in the AGN accretion disc. The motion of these masers takes place in the gravitational field of the BH characterized by the line element (\ref{schw}). Thus, one is able to observe the shift in the frequency of the photons that the masers emit at certain positions of their orbit.

In this context, the total frequency shift of the photons is given by 
\begin{equation}
1+z_{tot_{1,2}}=(1+z_{Schw_{1,2}}){(1+z_{rec}),}  \label{redshi_tot}
\end{equation}
where ${z_{rec}}$ is the recessional redshift composed by (Davis \& Scrimgeour \cite{Davis14})
{
\begin{equation}
1+z_{rec}=(1+z_{cosm})(1+z_{boost}), \qquad 1+z_{boost} = \frac{1+\beta}{\sqrt{1-\beta^2}}.
\end{equation}
In this relation, ${z}_{cosm}$ is the cosmological redshift, ${z_{boost}}$ is the frequency shift generated by a special relativistic boost that depends on the peculiar velocity $v_p$ characterizing the radial motion of the galaxy (Rindler \cite{RindlerSR1989}), and $\beta=v_p/c$. We defined the peculiar velocity in terms of the peculiar redshift as $v_p=c~z_p$. Besides, in Eq. (\ref{redshi_tot}) the Schwarzschild frequency shift, $z_{Schw_{1,2}}$, consists of the kinematic redshift and blueshift and the gravitational redshift as 
\begin{equation}
1+z_{Schw_{1,2}}=1+z_{grav}+{z_{kin_{\pm }},}  \label{redshi_scw}
\end{equation}%
in which $z_{grav}$ is the gravitational redshift due to the spacetime
curvature generated by the BH mass and $z_{kin_{\pm }}$ is the
kinematic redshift and blueshift of 
photons emitted by moving masers, that is, the local Doppler effect, with the following explicit form  
{
\begin{equation}
z_{grav}=\sqrt{\frac{1}{1-3\frac{m}{r_{e}}}}-1,\qquad z_{kin_{\pm }}=\pm 
\sqrt{\frac{\frac{m}{r_{e}}}{(1-3\frac{m}{r_{e}})(1-2\frac{m}{r_{e}})}},
\label{zg-zkin}
\end{equation}%
where $r_{e}$\ is the emitter radius (the radius of megamaser features in
this case). 
We used the approximation $ \Theta \approx r_e / D$, where $\Theta$ is the angular distance between a given maser and the BH in our model and estimations. 
In order to further develop our treatment, we needed to perform a statistical fit of megamaser astrophysical data. We remark that in this work, we focus on the redshifted and blueshifted masers only because they are located at the points where their velocity gain paths are the longest (i.e., around the disk midline).

\begin{table*}
\caption{
MCP reported parameters of the sample galaxies.}
\label{tab:prev}
\begin{center}
\footnotesize
\begin{tabular}{cccccccc}
Source & Position & Mass & Distance & $M/D$ & Recession velocity & Inclination angle &
\\ 
& R.A. ($^h:^m :^s$) &  &  &  &  &  & 
\\ 
& Decl. ($\degr: ^{^{\prime }} : ^{^{\prime \prime }}$) & ($10^7 M_{\odot}$)
& (Mpc) & ($10^5 M_{\odot}$)/Mpc & (km/s) & (\degr) & 
\\ \hline\hline \,
\multirow{2}{*}{NGC 1194$^e$} & 03:03:49.10864 & \multirow{2}{*}{6.5 $\pm$ 0.3}
& \multirow{2}{*}{55.4 $\pm$ 3.9*} & \multirow{2}{*}{11.733} & %
\multirow{2}{*}{4051 $\pm$ 15} & \multirow{2}{*}{85$^a$} & 
\\ 
& -01:06:13.4743 &  &  &  &  &  &  
\\ \hline
\multirow{2}{*}{NGC 5495$^f$} & 14:12:23.35 & \multirow{2}{*}{1.1$\pm$0.2} & %
\multirow{2}{*}{$95.7\pm5.3$ } & \multirow{2}{*}{ 1.149} & %
\multirow{2}{*}{6839.6 $\pm $67.0} & \multirow{2}{*}{$95\pm1$$^b$} & 
\\ 
& -27:06 29.20 &  &  &  &  &  &  
\\ \hline
\multirow{2}{*}{Mrk 1029$^f$} & 02:17:03.566 & \multirow{2}{*}{0.19 $\pm$ 0.05}
& \multirow{2}{*}{120.8 $\pm$ 6.6 } & \multirow{2}{*}{0.158} & %
\multirow{2}{*}{9129.9 $\pm$ 26.0} & \multirow{2}{*}{79 $\pm$2$^b$ } & 
\\ 
& 05:17:31.43 &  &  &  &  &  &  
\\ \hline
\multirow{2}{*}{NGC 1320$^f$} & 03:24:48.70 & \multirow{2}{*}{$0.55\pm 0.25$} & %
\multirow{2}{*}{$34.2\pm 1.9$ } & \multirow{2}{*}{1.608} & %
\multirow{2}{*}{2829.2 $\pm$ 11.4} & \multirow{2}{*}{90$^c$ } & 
\\ 
& -03:02:32.30 &  &  &  &  &  &  
\\ \hline
\end{tabular}%
\end{center}
\small \textbf{Notes.} Column 1: Name of the source. Column 2: Source position (J2000). Column 3: Fitted mass of the SMBHs. Column 4: Estimated distance to the sources. Column 5: Computed M/D ratio from Columns 3 and 4. Column 6: Fitted recession velocities. Column 7: Reported inclination angle $\mathbf{\theta_0}$ for NGC 1194, NGC 5495, and Mrk 1029 and fixed inclination angle for NGC 1320.\\
Label $^a$ represents values reported without uncertainty.\\
Label $^b$ refers to the inclination angle fixed through observations of systemic masers in Gao et al. (\cite{MCPIX}).\\
Label $^c$ refers to the fixed inclination angle due to the absence of systemic maser features in Gao et al. (\cite{MCPIX}).\\
Label $^d$ refers to the inclination angle fixed through observations of systemic masers in Zhao et al. (\cite{MCPX}).
\\ 
Label $^{*}$ indicates the distance to NGC 1194 estimated in Kamali et al. (\cite{ngc1194distance}).\\
Label $^e$ corresponds to the reference: Kuo et al. (\cite{MCPIII}).\\
Label $^f$ corresponds to the reference: Gao et al. (\cite{MCPIX}).
\\
\end{table*}

\section{Statistical fit}
\label{sec:bayes}

Before describing our modelling, we note that not all the detected masers lie perfectly along the midline; instead, they are spread in the vicinity of it. In order to take into account this scattering, we followed Herrnstein et al. (\cite{Herrnstein05}) and introduced into our model a small dispersion azimuthal angle $\delta\varphi$ responsible for the distribution of the masers close to the midline of the disk. We fixed this scattering angle by scanning its reasonable possible values in statistical estimations and choosing the smallest one that renders a reduced $\chi ^{2}$ close to unity.

That being said, in this paper, we applied a Bayesian statistical treatment to
observational data based on the Monte Carlo method that uses Markov chains. This treatment
consists of a least squares $\chi ^{2}$ fit of the $M/D$ ratio,
$z_{rec}$, and the displacement of both $x-$ and $y-$offsets of the central
object, which is given by
\begin{equation}
\chi ^{2}=\sum_{k=1}\frac{\left[ {z_{k,obs}}-(1+z_{grav}+\epsilon {\sin {%
\theta _{0}}}~z_{kin_{\pm }}){(1+z_{rec})}+1\right] ^{2}}{\sigma
_{z_{tot_{1,2}}}^{2}+{\kappa }^{2} {\epsilon^2} z_{kin_{\pm }}^{2}\sin ^{2}{\theta _{0}}{%
(1+z_{rec})^{2}}},  \label{chi}
\end{equation}%
where $\sigma _{z_{tot1,2}}=|\delta z_{tot_{1,2}}|$ is the variation of total redshift, $\theta_0$ is the inclination angle, and the quantities $\epsilon$ and ${ {\kappa}}$ refer to the spread of the redshifted and blueshifted features on the azimuthal angle:
 {\small    \begin{equation}
        \epsilon \approx 1 - \frac{\delta \varphi^2}{2} + \frac{\delta \varphi^4}{24}, \quad \quad { {\kappa}}^2 \approx \frac{\delta \varphi^4}{4}.
    \end{equation}}
Here, the first expansion corresponds to the cosine function of $\varphi$, and ${\kappa}$ denotes the induced uncertainties of the maser scattering under the assumptions $\varphi \ll 1$ and $ \varphi \sim \delta \varphi$. In addition, $\sigma
_{z_{tot_{1,2}}}^{2}$ in the $\chi ^{2}$ relation (\ref{chi}) represents the error of the total
redshift with the following explicit form:
 {\small 
\begin{equation}
\delta z_{tot_{1,2}}=(\delta z_{grav}+\delta z_{kin_{\pm }})(1+z_{rec}),
\end{equation}%
}where
\begin{equation}
{\small \delta z_{grav}=\left( 1+z_{grav}\right) ^{3}\left( \dfrac{-3m}{%
2r_{e}^{2}}\right) \delta r_{e},}
\end{equation}%
\begin{equation}
{\small \delta z_{kin_{\pm }}=\epsilon \sin {\theta _{0}}\left( z_{kin_{\pm
}}\right) ^{3}\left( \dfrac{6m^{2}-r_{e}^{2}}{2mr_{e}^{2}}\right) \delta r_{e},}
\end{equation}%
\begin{equation}
{\small \delta r_{e}=\sqrt{\left( \frac{x_{i}-x_{0}}{r}\right) ^{2}\delta
_{x}^{2}+\left( \frac{y_{i}-y_{0}}{r}\right) ^{2}\delta _{y}^{2}},}
\end{equation}
such that ($x_i$, $y_i)$ is the position of the $i$-th megamaser on the sky, $\{\delta _{x},\delta _{y}\}$ are their respective errors, and $(x_{0},y_{0})$ represents the BH position.

\subsection{Parameter fitting: Priors and posteriors}

The convention used by the MCP to allocate the origin of the reference system on the sky is to consider the brightest maser for NGC 5495, Mrk 1029, and NGC 1320  (see Gao et al. \cite{MCPIX}), whereas for NGC 1194, Kuo et al. (\cite{MCPIII}) chose the centroid of the redshifted and blueshifted masers for $y=0$, and the centroid of the systemic masers for $x=0$. We followed these works for NGC 1194, NGC 5495, and Mrk 1029. For NGC 1320, since systemic maser lines were not detected, we instead chose the reference position at the middle point between the geometrical centers of the highly frequency-shifted masers. Therefore, the BH offsets were calculated with respect to the latter referred positions.

In this paper, although
we used the database of Kuo et al. (\cite{MCPIII}) and Gao et al. (\cite{MCPIX}), we employed a rotated  distribution of the masers to the horizontal axis. 
We applied this rotation by considering the reported position angle (PA) for the megamaser systems.

This statistical fit takes into account the positions in the plane
perpendicular to the line of sight (LOS) of the 
redshifted and blueshifted water masers, with their
respective uncertainties in both the positions and the frequency shift of the
photons they emit.

The BH parameters that underwent the Bayesian statistical fit were the
mass-to-distance ratio $M/D$, the recessional redshift $z_{rec}$, and
the horizontal offset $x_{0}$ and the vertical offset $y_{0}$ of the BH
position. However, due to the characteristics of the maser distribution of individual galaxies,
the model did not render the estimations of all the parameters for each case. For instance, for most of the fitted galaxies, the parameter $y_{0}$ could not be estimated due to the thin character of the disk. 

Then, for the galaxies NGC 1194, Mrk 1029, and NGC 1320, we fixed the scattering angle of maser features $\delta \varphi$ as indicated above.
For NGC 5495, the $\chi ^{2}_{red}$ value was less than one. In this case, the maser position errors were much larger than those shown in the rest of galaxies under consideration (see Table \ref{tab:Results4par} and the corresponding Fig. \ref{fig:posteriors1}).

Finally, we calculated the gravitational redshift of the closest maser to the BH for these galaxies, and we present the results in Table \ref{tab:zgrav}. The contribution of this gravitational redshift in our estimations is an effect that has been properly included in our general relativistic study.

\begin{table*}
\caption{Best fitted values of our estimations.}
\label{tab:Results4par}
\begin{center}
\begin{tabular}{cccccccc}
Source & $M/D$ & $x_0$ & $y_0$ & $z_{rec}$ & $v_{rec}$ & $\delta \varphi$ & $%
\chi^2_{red}$ \\ 
& ($10^5 M_{\odot}/$Mpc) & (mas) & (mas) & ($10^{-2}$) & (km/s) & (%
${{}^\circ}$%
) &  \\ \hline\hline
\multirow{2}{*}{NGC 1194} & \multirow{2}{*}{$13.100^{+0.211}_{-0.209}$} & %
\multirow{2}{*}{$0.379 ^{+0.133}_{-0.131}$} & \multirow{2}{*}{$0.0$*} & %
\multirow{2}{*}{$1.363 \pm 0.004$} & %
\multirow{2}{*}{$4085.984^{+12.527}_{-12.421}$} & \multirow{2}{*}{11} & %
\multirow{2}{*}{1.206} \\ 
&  &  &  &  &  &  &  \\ \hline
\multirow{2}{*}{NGC 5495} & \multirow{2}{*}{$1.153^{+0.208}_{-0.174}$} & %
\multirow{2}{*}{$-0.071^{+0.174}_{-0.118}$} & \multirow{2}{*}{$0.0$*} & %
\multirow{2}{*}{$2.281^{+0.018}_{-0.019}$} & %
\multirow{2}{*}{$6839.055^{+53.393}_{-57.772}$} & \multirow{2}{*}{0} & %
\multirow{2}{*}{0.282} \\ 
&  &  &  &  &  &  &  \\ \hline
\multirow{2}{*}{Mrk 1029} & \multirow{2}{*}{$0.144\pm 0.011$} & %
\multirow{2}{*}{$-0.376^{+0.143}_{-0.138}$} & \multirow{2}{*}{$0.0$*} & %
\multirow{2}{*}{$3.035 \pm 0.004$} & %
\multirow{2}{*}{$9099.332^{+12.710}_{-11.252}$} & \multirow{2}{*}{14} & %
\multirow{2}{*}{1.385} \\ 
&  &  &  &  &  &  &  \\ \hline\hline
\multirow{2}{*}{NGC 1320} & \multirow{2}{*}{$1.497^{+0.071}_{-0.069}$} & %
\multirow{2}{*}{$-1.243^{+0.147}_{-0.132}$} & %
\multirow{2}{*}{$-0.761^{+0.149}_{-0.142}$} & \multirow{2}{*}{$0.942 \pm
0.002$} & \multirow{2}{*}{$2825.433^{+6.158}_{-6.389}$} & \multirow{2}{*}{17}
& \multirow{2}{*}{1.127} \\ 
&  &  &  &  &  &  &  \\ \hline
\end{tabular}%
\end{center}
\small \textbf{Notes.} Column 1: Name of the source. Column 2: Fitted mass-to-distance ratio. Column 3: Horizontal offset ($x_0$) of the BH found by the Bayesian fit. Column 4: Vertical offset ($y_0$) of the BH either found by the Bayesian fit or fixed. Column 5: Recessional redshift found by the fit. Column 6: Recession velocity after the optical definition of the redshift. Column 7: Scattering in the azimuth angle. Column 8: Reduced $\protect%
\chi^2$ of the best fit.
The double horizontal line separates different galaxies according to the number of estimated parameters. We estimate three parameters for the first three galaxies and four parameters for the fourth galaxy.\\
The label $*$ indicates that we fixed this parameter.
\end{table*}

\subsection{NGC 1194}

NGC 1194 is a system with three groups of maser features (see Fig. \ref{fig:posteriors1}). The information of this maser system is consistent with a thin disk that is distributed on the sky with a PA of approximately 157$\degr$. 
For this maser distribution, our reference position is calculated as the centroid of the redshifted and blueshifted masers for $y=0$ and the centroid of the systemic masers for $x=0$, as in Kuo et al. (\cite{MCPIII}).
We performed the statistical fit to the mass-to-distance ratio, the recessional redshift, and the BH $x$-offset, while we fixed its $y$-offset aligned with the redshifted and blueshifted masers. The estimated $ M/D =(1.31\pm 0.02)\times 10^{6}M_{\odot}/Mpc$ implies $ M = (7.26 \pm 0.52)\times 10^{7}M_{\odot}$ for the BH mass when taking into account the distance reported in Table \ref{tab:prev}. By considering the $v_{rec}$ estimation of the general relativistic approach and the reported recession velocity in Kuo et al. (\cite{MCPIII}; see Table \ref{tab:prev}), we found that these results exclude each other. 
It is worth mentioning that since the general relativistic model employed
in this paper is different from prior works, the source of this
disagreement and obtaining different results is due to the difference
in the modelling. 
The scattering angle $\delta \varphi=11\degr$ renders $\chi_{red}^{2}=1.206$.

\subsection{NGC 5495}

NGC 5495 is a Seyfert 2 AGN-type galaxy with an original north-south orientation maser disk and a PA of $176\degr\pm 5\degr$. 
As in Gao et al. (2016), we used the brightest systemic maser with velocity $6789.03$ km s$^{-1}$ as the reference position on the sky. 
For this galaxy, we estimated the three parameters: the ratio $M/D=1.15^{+0.21}_{-0.17}\times 10^{5}M_{\odot }/Mpc$ that implies a BH mass of $M=1.10^{+0.21}_{-0.18}\times 10^{7}M_{\odot}$, the recession velocity $v_{rec}$, and the BH $x$-offset with respect to the above-mentioned reference point. The fit was performed without a scattering angle, leading to a reduced $\chi _{red}^{2}=0.28$. The resulting BH position is optimal in the sense that the statistical fit renders it just behind the systemic masers (see Fig. \ref{fig:posteriors1}).
For this maser distribution, a 5$\degr$ error in the PA has been reported (Gao et al. \cite{MCPIX}). By propagating the error in the position angle during rotation, we obtained a 0.16\% change in the estimated M/D ratio.

\subsection{Mrk 1029}
Mrk 1029 is a galaxy with an original linear maser distribution from the north-east to the south-west at a PA of $218\degr\pm 10\degr$. 
Following Gao et al. (2016), we used the brightest systemic maser with velocity $9136.86$ km s$^{-1}$ as the reference position on the sky and considered a flat disk consisting of particles in circular motion. 
For this galaxy, we estimated three
parameters, namely, the ratio $M/D=(1.44 \pm 0.11)\times
10^{4}M_{\odot }/Mpc$ that implies a BH mass $M=(1.74\pm0.16)\times 10^{6}M_{\odot
}$, the recession velocity $v_{rec}$, and the $x$-offset of the BH position. These parameters were fitted for a scattering angle of $\delta \varphi =14\degr$ and led to a reduced $\chi ^{2}_{red}=1.385$. For this galaxy, the recession velocities of 9076 $\pm$ 32 km s$^{-1}$ and 9160 $\pm$ 61 km s$^{-1}$ reported in Huchra et al. (\cite{huchra99}) and De Vaucoleurs et al. (\cite{Vaucouleurs}), respectively, are compatible with our estimated value of $9099.33^{+12.71}_{-11.25}$ km s$^{-1}$ (see Table \ref{tab:Results4par}). Again, the resulting BH position is optimal in the sense that the fit yields its location among the systemic masers.

On the other hand, by propagating the PA error during the rotation of the maser system, the $M/D$ estimation  changes by about 1.19\%, a quantity that is within its error (see Table \ref{tab:Results4par}). Hence, we could neglect the error associated with the PA.

\subsection{NGC 1320}

The masers of NGC 1320 have been found in just two distinct frequency-shifted groups. The maser distribution shows an east-west orientation with a PA of 
$75\degr \pm 10 \degr$. 
Since systemic maser lines were not detected, we chose the reference position at the middle point between the geometrical centers of the highly frequency-shifted masers.
For this system, we estimated four parameters: 
the mass-to-distance ratio $M/D=1.50\pm 0.07\times10^{5}M_{\odot }/Mpc$, the recession velocity, and the BH $x-$ and $y-$offsets with respect to the above-referenced position. 
In this case, the corresponding BH mass reads as $M=(5.12\pm0.37)\times 10^{6}M_{\odot}$. By considering $\delta \varphi=17\degr$,  the model gives a reduced $\chi^2_{red}=1.127$.

With this general relativistic  model, the $x$-offset of the BH led to a position very close to the blueshifted maser group. 
In addition, by considering the propagation of the PA error under the rotation, we obtained a slight difference of 2.64\% in the estimation of the $M/D$ ratio. Despite being the galaxy with the biggest change in the mass-to-distance ratio, it is still within the $M/D$ error,  as can be seen from Table \ref{tab:Results4par}.

\begin{figure*}
	\includegraphics[scale=.842]{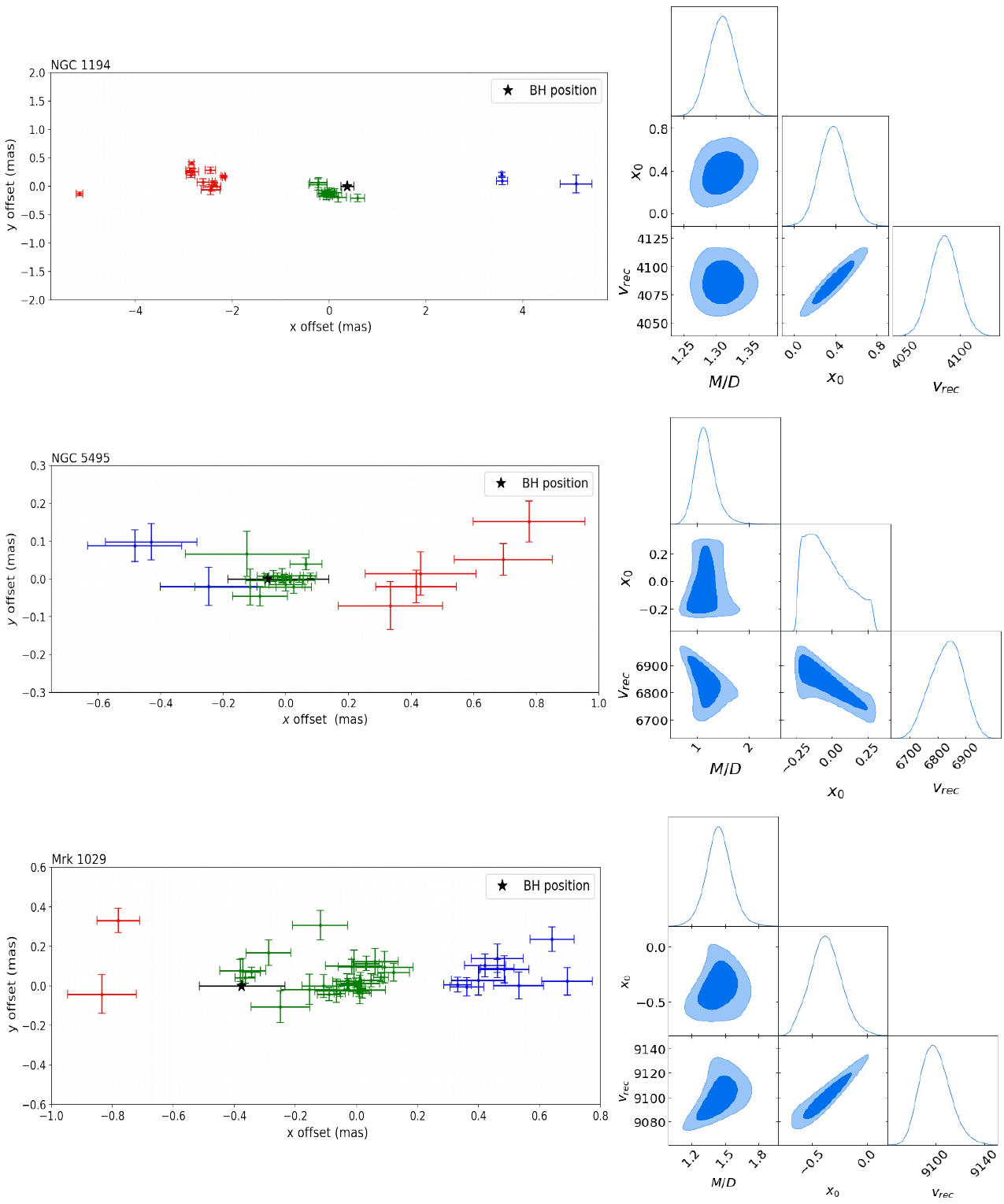}
    \caption{Maser disks and posterior probability distributions for the galaxies NGC 1194, NGC 5495, and Mrk 1029. The maser disk plots show the three groups of maser features viewed edge-on with their observational error. The star symbol refers to the best fit for the BH position on the disk. The black line through this star symbol indicates the uncertainty in the position of BH. The blue graphs in the right panels show the posterior probability distribution with the contour levels corresponding to 1$\sigma$ and 2$\sigma$ confidence regions.}
    \label{fig:posteriors1}
\end{figure*}

\begin{figure*}
	\includegraphics[scale=.842]{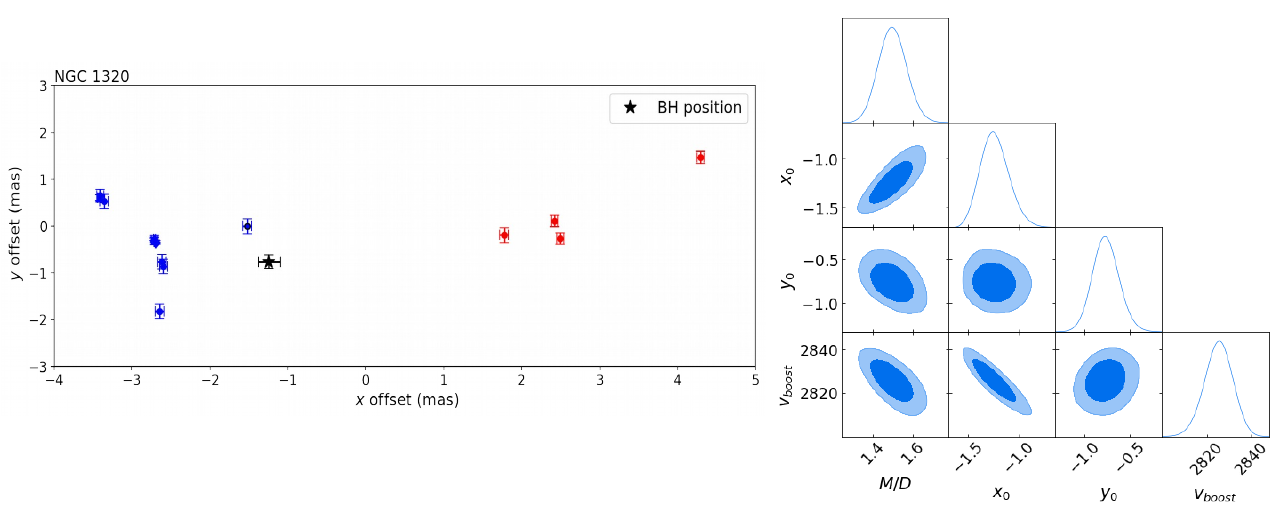}
    \caption{Maser disk and posterior probability distribution for the galaxy NGC 1320. The maser disk plot shows an AGN with particular characteristics. NGC 1320 does not have a systemic maser feature group, and the model allows for the estimation of four parameters. The star symbol in the left panel indicates the best fit for the BH position on the disk, and the black line through this symbol denotes the uncertainty in the BH position. The blue graph in the right panel shows the posterior probability distribution, with the contour levels corresponding to 1$\sigma$ and 2$\sigma$ confidence regions.}
    \label{fig:posteriors2}
\end{figure*}

\begin{table}
\caption{Gravitational redshift of closest masers to BHs.}
\label{tab:zgrav}
\begin{center}
\footnotesize
\begin{tabular}{ccccc}
& Maser & Distance to the & Gravitational & Associated \\ 
Source & feature & closest maser & redshift & velocity \\ 
& & (mas) & ($10^{-6}$) & (km/s) \\ \hline\hline
NGC 1194 & Red & 2.549 & 7.610 & 2.282 \\ 
NGC 5495 & Blue & 0.189 & 9.036 & 2.709 \\ 
Mrk 1029 & Red & 0.460 & 0.465 & 0.139 \\ 
NGC 1320 & Blue & 0.809 & 2.740 & 0.821 \\
\hline
\end{tabular}%
\end{center}
\end{table}

\section{Discussion and conclusions} \label{sec:conclusions}

In this work, the four AGNs NGC 1194, UGC 5495, Mrk 1029, and NGC1320 from the MCP have been studied in order to fit the $M/D$ ratio of the central BH hosted at their cores, along with other parameters such as 
recession velocity and 
position on the sky. These results have been obtained by employing a full general relativistic method that allowed us to clearly identify the relativistic effects involved in the dynamics of the above-mentioned megamaser systems. In particular, it provided an analytic expression for the gravitational redshift that depends on the BH mass and the orbital radius of the test particles.

It is worth mentioning that since we estimated the M/D ratio of several BHs in this paper by using a general relativistic method, our results are different from those of the MCP, where the corresponding BH masses and distances were estimated through a Newtonian approach with relativistic corrections. Therefore, we cannot compare both results directly since different quantities were estimated. 
Notwithstanding, we are currently developing a method to decouple $M$ and $D$ so that the results of each approach can be directly compared in the near future.

In addition, we have observed that, in general, the recession velocities of these galaxies are compatible with the corresponding estimations reported in Table \ref{tab:prev}, except for NGC 1194. For this galaxy, the recession velocities yielded by both fits exclude each 
other: The result reported by Kuo et al. (\cite{MCPIII}) is $4051 \pm 15$ \text{km s}$^{-1}$, whereas our best estimated value is 
$4085.98 \pm 12.53$ \text{km s}$^{-1}$. However, in a more recent study (Pesce et al. \cite{Pesce2018}), the authors fitted the same megamaser (observational) data with a slightly different rotation curve and found a recession velocity of $4088.8 \pm 5.3$ \text{km s}$^{-1}$. Moreover, these authors obtained another estimate of $4088.6 \pm 5.8$ \text{km s}$^{-1}$ from an H{\small I} tilted-ring fitting. Both of these results are in agreement with the value of the recession velocity  obtained in our work.

One more peculiarity regarding our results consists of the best-estimated value that we obtained for the BH position in NGC 1320, which is very close to the blueshifted maser group (see Fig. 2), in agreement with Gao et al. (\cite{MCPIX}). Indeed, Gao et al. reported that the maser profile in this galaxy is not typical since it possesses just two different velocity groups, it is highly variable, and the corresponding VLBI map suggests a strong disk warp or a maser emission different from the disk origin. Thus, by taking advantage of the high variability of the masers in NGC 1320, one could perform new VLBI measurements that would allow for better characterization of both the maser system and its BH position.

On the other hand, the general relativistic formalism that we have employed to estimate the BH parameters allowed us to quantify the gravitational redshift of each highly frequency-shifted maser. 
In this regard, we have calculated for the first time the gravitational redshift of the closest masers to each SMBH hosted at the core of the studied astrophysical AGNs.
The results are presented in Table \ref{tab:zgrav}. In principle, the magnitude of the gravitational redshifts of these masers is compatible with the magnitude of the gravitational redshifts that has been computed for a dozen megamaser systems studied in Nucamendi et al. (\cite{Nucamendi21}), Villalobos-Ram\'\i rez et al. (\cite{Artemisa22}), and Villaraos et al. (\cite{Deby1}). We emphasize that all of these gravitational redshift computations constitute the first applications of our general relativistic method to real astrophysical systems based on analytic predictions, a fact that in principle facilitates its potential detection. 

Moreover, in order to compare the M/D ratio presented in this study with the MCP estimations of BH mass, we calculated the mass-to-distance ratio using the MCP results by dividing the mass over the distance 
(see Column 5 of Table \ref{tab:prev}). 
We did not propagate the mass and distance uncertainties in the $M/D$ ratio since the uncertainty in the mass-to-distance ratio of the MCP results inherits the large distance error during propagation, and thus the resultant uncertainties would be misleading. 
On the other hand, we have calculated the BH mass from the M/D estimations multiplied by the distance taken from Table \ref{tab:prev} and propagated the corresponding uncertainty for each galaxy.

Finally, we finish our paper with a couple of suggestions for future work. Within this general relativistic model, considering the systemic masers in the statistical fits in order to improve the accuracy of the estimations would be interesting. Moreover, it would be worthwhile to break the M/D degeneracy in order to obtain independent estimations for the BH mass and the distance to the BH. These extensions are currently under examination.

\begin{acknowledgements}
All authors are grateful to D. Villaraos for shared experience and illuminating discussions about previous work, to A. Christen for sharing his expertise on statistical estimations and to an anonymous referee for insightful and constructive observations that improved the quality of our work. 
The authors thank the Megamaser Cosmology Project researchers for making the observational data used in this work publicly available and CONACYT for support under grant No. CF-MG-2558591. A.G.-J. and M.M. thank SNII and were supported by CONACYT through the postdoctoral grants Nos. 446473 and  31155, respectively. A.V.-R acknowledges financial assistance from CONACYT through grant No. 1007718.
A.H.-A. was supported by VIEP-BUAP as well as by SNII.
\end{acknowledgements}

%
%

\end{document}